\documentclass[aps,prl,twocolumn,showpacs,groupedaddress,nofootinbib]{revtex4}
\usepackage{graphicx}  
\usepackage{dcolumn}   
\usepackage{bm}        
\usepackage[english]{babel}
\usepackage{amsfonts,amsmath,amssymb,mathrsfs}
\usepackage{times}

\usepackage{latexsym}

\newcommand{\eq}{\begin{equation}}
\newcommand{\eeq}{\end{equation}}

\newcommand{\be}{\begin{equation}}
\newcommand{\ee}{\end{equation}}
\newcommand{\bea}{\begin{eqnarray}}
\newcommand{\nn}{\nonumber}
\newcommand{\eea}{\end{eqnarray}}

\def\two{\hbox{$_{2}$}}
\def\three{\hbox{$_{3}$}}
\def\four{\hbox{$_{4}$}}

\begin{document}

\title{Post-Newtonian expansion for Gauss-Bonnet Gravity}

\author{Thomas Sotiriou}\email{sotiriou@sissa.it}
\affiliation{SISSA, International School for
             Advanced Studies, Via Beirut 2, 34014 Trieste, Italy and INFN, Sezione di Trieste}
\author{Enrico Barausse}\email{barausse@sissa.it}
\affiliation{SISSA, International School for
             Advanced Studies, Via Beirut 2, 34014 Trieste, Italy and INFN, Sezione di Trieste}

\date{\today}
\begin{abstract}
The Parametrized Post-Newtonian expansion of gravitational theories with a scalar field coupled to the Gauss-Bonnet invariant is performed and confrontation of such theories with Solar system experiments is discussed.
\end{abstract}

\maketitle
The observed late time accelerated expansion of the universe \cite{riess} combined with the problems that arise when trying to rationalize it with the simplest of explanations, such as a cosmological constant \cite{carroll}, have triggered an increased interest in finding alternatives for the nature of dark energy. Scalar fields, widely used in the inflationary paradigm \cite{guth}, constitute a familiar way of providing accelerated expansion. Scalar-tensor theory is therefore an appealing candidate as an alternative theory of gravity that can provide the desired cosmological dynamics \cite{valerio}. However, there are motivations from String theory to believe that scalar fields might not be coupled to the Ricci scalar, as in scalar-tensor theory, but to the Gauss-Bonnet term, 
\be
{\cal G}=R^2-4 R^{\mu\nu}R_{\mu\nu}+R^{\mu\nu\kappa\lambda}R_{\mu\nu\kappa\lambda},
\ee
which is topologically invariant in four dimensions.

To be more precise, one expects to find two types of scalar fields in the low energy effective action of gravity coming from heterotic String theory: moduli, $\phi$, which are related to the size and shape of the internal compactification manifold, and the dilaton $\sigma$, which plays the role of the string loop expansion parameter. There are reasons to believe that moduli generally couple to curvature square terms \cite{strings1} but moduli dependent higher loop contributions, such as terms cubic or higher order in the Riemann tensor, vanish, leaving a coupling with a Gauss-Bonnet term to be of specific interest \cite{strings1,strings2}. On the other hand the dilaton usually couples to the the Ricci scalar and consequently to matter in the Einstein frame, and there are reasons to believe that it might evolve in such a way as to settle to a constant \cite{strings3}. Under these assumptions the effective low energy gravitational action takes the form
\be
\label{action}
S=\int d^4 x \sqrt{-g}\left[\frac{R}{2 \kappa^2}-\frac{\lambda}{2}\partial_\mu \phi \partial^\mu \phi-V(\phi)+f(\phi) {\cal G}\right],
\ee
where $\kappa^2=8\pi G$, $\lambda$ is $+1$ for a canonical scalar and $-1$ for a phantom field ($c=\hbar=1$). A straightforward generalization of the action is to include a kinetic term and a coupling with ${\cal G}$ for the dilaton $\sigma$. This will not concern us here but we will discuss how we expect it to affect our results.

Remarkably, it has been shown that action (\ref{action}) can lead to a theory of gravity with desirable phenomenology, including inflation and accelerated expansion \cite{nojodsas,cosm}. Such a theory seems to fit observational data related to cosmology adequately \cite{tomi} and avoid past and future singularities \cite{strings2, nojodsas}. However, a gravitational theory which can pose a viable alternative to General Relativity should also have the correct Newtonian and post-Newtonian limits, since Solar system tests provide stringent constraints. This is the issue that will concern us here. We will not consider the case where $\lambda=0$ and the scalar field has no kinetic term in the action. Such actions are dynamically equivalent to an action with a general function of ${\cal G}$ added to the Ricci scalar and their Newtonian limit has already been considered in \cite{odi}.


 Let us start by reviewing the field equations one derives from action (\ref{action})~\cite{nojodsas}. For the metric we have
\bea
\label{gfield}
G_{\mu\nu}&=&\kappa^2\Big[T_{\mu\nu}+T_{\mu\nu}^\phi+2(\nabla_\mu \nabla_\nu f(\phi)) R-\\& &-2 g_{\mu\nu}(\Box f(\phi)) R-4(\nabla^\rho \nabla_{\mu} f(\phi))R_{\nu\rho}-\nn\\& &-4(\nabla^\rho \nabla_{\nu} f(\phi))R_{\mu\rho}+4(\Box f(\phi))R_{\mu\nu}+\nn\\& &+4g_{\mu\nu}(\nabla^\rho \nabla^\sigma f(\phi))R_{\rho\sigma}-4(\nabla^\rho \nabla^\sigma f(\phi))R_{\mu\rho\nu\sigma}\Big]\nn,
\eea
where $G_{\mu\nu}\equiv R_{\mu\nu}-\frac{1}{2}R g_{\mu\nu}$ and
\be
\label{tphi}
T_{\mu\nu}^\phi=\lambda\left(\frac{1}{2}\partial_\mu \phi \partial_\nu \phi-\frac{1}{4}g_{\mu\nu}\partial^\rho \phi \partial_\rho \phi\right)-\frac{1}{2}g_{\mu\nu}V(\phi),
\ee
and for the scalar field
\be
\label{phifield}
\lambda \Box \phi-V'(\phi)+f'(\phi){\cal G} =0,
\ee
where $A'(x)\equiv \partial A/\partial x$ and  $\Box\equiv g^{\mu\nu}\nabla_\nu\nabla_\mu$.

Let us bring eq.~(\ref{gfield}) to a more suitable form for our purposes. Taking the trace of eq.~(\ref{gfield}) one gets
\be
\label{trace}
R=\kappa^2 \left[-T-T^\phi+2 (\Box f(\phi))R-4(\nabla^\rho\nabla^\sigma f(\phi))R_{\rho\sigma}\right],
\ee
where $T=g^{\mu\nu}T_{\mu\nu}$ and $T^\phi=g^{\mu\nu}T_{\mu\nu}^\phi$.
Replacing eq.~(\ref{trace}) back in eq.~(\ref{gfield}), the latter becomes:
\bea
\label{gfield2}
R_{\mu\nu}&=&\kappa^2\Big[T_{\mu\nu}-\frac{1}{2}g_{\mu\nu}T+\frac{1}{2}\lambda\partial_\mu \phi \partial_\nu \phi+\frac{1}{2}g_{\mu\nu}V(\phi)+\nn\\& &+2(\nabla_\mu \nabla_\nu f(\phi)) R-g_{\mu\nu}(\Box f(\phi)) R-\nn\\& &-4(\nabla^\rho \nabla_{\mu} f(\phi))R_{\nu\rho}-4(\nabla^\rho \nabla_{\nu} f(\phi))R_{\mu\rho}+\nn\\& &+4(\Box f(\phi))R_{\mu\nu}+2g_{\mu\nu}(\nabla^\rho \nabla^\sigma f(\phi))R_{\rho\sigma}-\nn\\& &-4(\nabla^\rho \nabla^\sigma f(\phi))R_{\mu\rho\nu\sigma}\Big]
\eea


Following \cite{will} we can choose a system of coordinates in which the metric can be perturbatively expanded around Minkowski spacetime. Therefore we write the metric as $g_{\mu\nu}=\eta_{\mu\nu}+h_{\mu\nu}$ and the scalar field as $\phi=\phi_{0}+\delta \phi$,
where the value of $\phi_0$ is determined by the cosmological solution. 
The perturbed field equations are
\begin{widetext}
\begin{eqnarray}
&&\lambda[\Box_{\rm flat}\delta\phi+(\delta\Box)\delta\phi]-V^{\prime\prime}(\phi_0)\delta \phi-\frac12 V^{\prime\prime\prime}(\phi_0)(\delta \phi)^2+f^\prime(\phi_0) {\cal G}={\cal O}(\delta\phi^3,\delta\phi(h_{\mu\nu})^2,h_{\mu\nu}\dot\phi_0,h_{\mu\nu}\ddot\phi_0),\\\label{00_perturbed_eq}
R_{00}&=&\kappa^2\Big\{T_{00}+\frac12T-\frac12h_{00}T+\frac{1}{2}\lambda\partial_0 \delta\phi \partial_0 \delta\phi
+\frac12\lambda\,\dot\phi_0^2-\frac12V(\phi_0)+\frac12V'(\phi_0)\delta\phi\, (-1+h_{00})+\nonumber\\
&&+f'(\phi_0)\Big[2(\partial_0\partial_0 \delta\phi) R+(\Box_{\rm flat} \delta\phi) R-8(\partial^\rho \partial_0 \delta\phi)R_{0\rho}
+4(\Box_{\rm flat}\delta\phi)R_{00}-2(\partial^\rho \partial^\sigma \delta\phi)R_{\rho\sigma}-\nonumber\\&&-4(\partial^\rho \partial^\sigma \delta\phi)R_{0\rho0\sigma}\Big]\Big\}+{\cal O}(\delta\phi^2h_{\mu\nu},\delta\phi^3,\dot\phi_0\delta\phi,\ddot\phi_0h_{\mu\nu},V(\phi_0)h_{00}),\\
R_{0i}&=&\kappa^2 T_{0i}+{\cal O}(\delta\phi h_{\mu\nu},\delta\phi^2,Th_{0i},\dot\phi_0\delta\phi,\ddot\phi_0h_{\mu\nu},V(\phi_0)h_{0i}),\\
R_{ij}&=&\kappa^2\left[T_{ij}+\frac12\delta_{ij}\left(-T+V'(\phi_0)\delta\phi+V(\phi_0)\right)\right]+{\cal O}(\delta\phi h_{\mu\nu},\delta\phi^2,Th_{ij},\ddot\phi_0h_{\mu\nu},V(\phi_0)h_{ij}),
\end{eqnarray}
\end{widetext}
where $\Box_{\rm flat}$ denotes the D'Alembertian of flat spacetime. Notice that, as usually done in scalar-tensor theory \cite{wagoner,pwill}, we have neglected all of the terms containing derivatives of $\phi_0$ multiplying perturbed quantities (e.g. $\dot\phi_0 \delta\phi$). This is due to the fact that  $\phi_0$ changes on cosmological timescales and consequently one expects that it remains practically constant during local experiments. Therefore its time derivatives can be neglected as far as solar system tests are concerned. 
This can easily be verified by some order-of-magnitude analysis. Take for instance Eq.~(\ref{00_perturbed_eq}): the terms containing a time derivative of $\phi_0$ multiplying a perturbation are ${\cal O}(\ddot{f}(\phi_0)h_{\mu\nu}/(r^2M_p^2))$ and ${\cal O}(\dot\phi_0\delta\dot\phi/M_p^2)$, where  $\dot\phi_0\sim H_0M_p$ and $\ddot{f}\sim H_0^2$ ($M_p=1/\kappa$ is the Planck mass and $H_0$ the present Hubble constant) and $h_{00}\sim h_{ij}\sim r\delta\phi\sim h_{0i}/v\sim r^2 \delta\dot\phi/v\sim GM_{\odot}/r$ ($r$ is the distance from the Sun and $v=\sqrt{GM_{\odot}/r}$).
On the other hand, the ${\cal O} (v^4)$ post-Newtonian correction to  $R_{00}$ is $\sim (GM_\odot)^2/r^4\sim 10^{-55}{\cal O}(\ddot{f}(\phi_0)h_{\mu\nu}/(r^2M_p^2), \dot\phi_0\delta\dot\phi/M_p^2)$ even if $r$ is taken as large as $1000$ AU. Therefore, the corrections coming from terms containing time derivatives of $\phi_0$ multiplying perturbations are at least 55 orders of magnitude smaller than post-Newtonian corrections, and neglecting these terms cannot affect our results in any way.
A similar treatment applies to the terms containing the potential $V$ multiplying perturbed quantities (e.g.$V(\phi_0)h_{00}$): in order to give a reasonable cosmology, $V(\phi_0)$ should be of the same order as the energy density of the cosmological constant and these terms cannot therefore  lead to any observable deviations at Solar system scales. 

We will instead adopt a different treatment required for the simple $V(\phi_0)$, $\frac12\dot\phi^2$ terms appearing in the field equations: as they need to be of the same order as the energy density of the cosmological constant, they will not lead to any observational consequences (see \cite{sereno} and references therein). However, for the sake of the argument we will keep track of them: due to their small value we can treat them as ${\cal O}(v^4)$ quantities following \cite{sereno} and so they will not appear in the ${\cal O} (v^2)$ equations. As far as terms related to $V'(\phi_0)$ are concerned, we intend to just keep track of them for the time being and discuss their contribution later on. 

If we now expand the perturbations in the metric and the scalar field in post-Newtonian orders, keeping in mind that the Parametrized Post-Newtonian expansion requires $\phi$ and $h_{00}$ to ${\cal O} (v^4)$, $h_{ij}$ to ${\cal O} (v^2)$ and $h_{0i}$ to ${\cal O} (v^3)$, we can write
\bea
\delta \phi&=&\two\delta \phi+\four\delta \phi\ldots\\
h_{00}&=&\two h_{00}+\four h_{00}\ldots\\
h_{ij}&=&\two h_{ij}+\ldots\\
h_{0i}&=&\three h_{0i}+\ldots
\eea
where the subscript denotes the order in the velocity, {\em i.e.~}quantities with a subscript $\two$ are ${\cal O}(v^2)$, quantities with a subscript $\three$ are ${\cal O}(v^3)$, etc. So to order ${\cal {O}}(v^2)$ this gives 
\be
\label{2phi}
\lambda\nabla^2(\two\delta\phi)-V^{\prime\prime}(\phi_0)\two\delta\phi=0\;:
\ee
where $\nabla^2\equiv\delta_{ij}\partial_i\partial_j$. Note that since the metric is flat in the background ${\cal G}={\cal O} (v^4)$. We want $\phi$ to take its cosmological value at distances far away from the sources. This is equivalent to saying that the perturbations due to the matter present in the Solar system should vanish at cosmological distances, and this can be achieved by imposing asymptotic flatness for the solution of eq.~(\ref{2phi}), {\em i.e.~}$\two\delta\phi \to 0$ for $r \to \infty$. This implies $\two\delta\phi=0$.

To order ${\cal O}(v^2)$ for the components $00$ and $ij$ and ${\cal O}(v^3)$ for the components $0i$, and after applying the  standard gauge conditions
\be
\label{gauge1}
h^\mu_{i,\mu}-\frac{1}{2}h^\mu_{\mu,i}=0\;,\qquad
h^\mu_{0,\mu}-\frac{1}{2}h^\mu_{\mu,0}=\frac12 h^0_{0,0}\;,
\ee
 the field equations for the metric take the form
\bea
-\nabla^2{(\two h_{00})}&=&\kappa^2\rho\\
-\nabla^2{(\two h_{ij})}&=&\kappa^2\rho\delta_{ij}\\
{1\over 2}\left(\nabla^2{(\three h_{0i})}+\frac12(\two h_{00,j0})\right)&=&\kappa^2\rho v^i
\eea
which remarkably are exactly the same as in General Relativity \cite{will}. The well-known solutions are
\bea
\two h_{00}&=&2U,\label{two_00}\\
\two h_{ij}&=&2U\delta_{ij}\label{two_ij},\\
\three h_{0i}&=&-\frac{7}{2}V_i-\frac{1}{2}W_i
\eea
where following \cite{will} we define
\bea
U&=&G\int d^3 x'\frac{\rho(x',t)}{|x-x'|},\\
V_i&=&G\int d^3 x'\frac{\rho(x',t) v_i(x',t)}{|x-x'|},\\
W_i&=&G\int d^3 x'\frac{\rho(x',t) v^k(x',t)(x-x')_k(x-x')_i}{|x-x'|^3}.
\eea

We already see that the theory has no deviation from General Relativity at order ${\cal O} (v^3)$: in particular it gives the correct Newtonian limit. It is now easy to go one step further and write down the perturbed equations that we need to ${\cal O}(v^4)$. For the scalar field, using $\two\delta\phi=0$, we get
\begin{equation}
\label{4phi}
\lambda\nabla^2(\four\delta\phi)-V^{\prime\prime}(\phi_0)\, \four\delta\phi+f^\prime(\phi_0)\, \four {\cal G}=0\;,
\end{equation}
with
\bea
\label{gbpert}
\four {\cal G}&=&(\two h_{00,ij})(\two h_{00,ij})-(\two h_{00,ii})(\two h_{00,jj})+(\two h_{ij,ij})^2+\nn\\& &+(\two h_{ij,kl})(\two h_{ij,kl})-(\two h_{ij,kk})(\two h_{ij,kk})-\nn\\& &-2 (\two h_{ij,kl})(\two h_{il,jk})+(\two h_{ij,kl})(\two h_{kl,ij})\;,
\eea
where we have again applied the gauge conditions (\ref{gauge1}).
Using eqs.~(\ref{two_00}) and~(\ref{two_ij}),  eq.~(\ref{gbpert}) becomes
\be
\label{4G}
\four {\cal G}=8\,U_{,kl}U_{,kl}-8\,(U_{,kk})^2\;.
\ee
The solution of eq.~(\ref{4phi}) is therefore
\be
\label{phisol}
\four\delta\phi=\frac{f^\prime(\phi_0)}{4\,\pi}\,
\int d^3 x'\frac{\four {\cal G}(x',t)}{|x-x'|} e^{-\sqrt{V^{\prime\prime}(\phi_0)}|x-x'|}
\ee

The time-time component of the perturbed field equations for the metric to ${\cal O}(v^4)$ is
\bea
\label{4hoo}
\four R_{00}&=&\kappa^2\Big[(\four T_{00})+\frac12 (\four T)-\frac12(\two h_{00})(\two T)\nonumber\\
&&-\frac12V'(\phi_0)(\four\delta\phi)-\frac12V(\phi_0)+\frac12\lambda\dot\phi_0^2\Big],
\eea
where we have already used the fact that $\two\delta\phi=0$. Note also that no contribution coming from the coupling between $\phi$ and the curvature terms in eq.~(\ref{gfield}) is present in the above equations. This should have been expected since in eq.~(\ref{gfield}) these terms always have the structure of two derivatives of $\phi$ times a curvature term, and so, due to the fact that in the background the metric is flat and $\phi_0$ is slowly varying, they can only contribute to orders higher than ${\cal O}(v^4)$.

Let us discuss the contibution of the term proportional to $V'(\phi_0)$. Using eqs.~(\ref{phisol}) and (\ref{4G}) we can write this term as an integral over the sources times a dimensionless coefficient $\kappa^2 V'(\phi_0)f'(\phi_0)$. One can argue the $V'(\phi)$ should be practically zero as far as the post-Newtonian expansion is concerned \cite{wagoner,pwill}. This is equivalent to saying that the cosmological solution corresponds to a minimum of the potential. Even though such assumptions are not exact, they are accurate enough for our purposes. Note that even in cases where $V$ does not have a minimum, well motivated models usually introduce exponential forms for the potential and the coupling function, {\textit i.e.~}$V=V_0 e^{-a\kappa\phi}$ and $f=f_0 e^{b\kappa\phi}$ where $a$, $b$ and $f_0$ are of order unity whereas $V_0$ is as small as the energy density of the cosmological constant in order to guarantee that the theory will fit observations related to the late time cosmological expansion. This implies that, since $\kappa^2\sim 1/M_p^2$, $\kappa^2 V'(\phi_0)f'(\phi_0)$ is dimensionless and of the order of the now renowned $10^{-123}$. Therefore, we will not take the term proportional to $V'(\phi_0)$ into account for what comes next but we will return to this discussion shortly.

 We can use the solutions for $\two h_{00}$ and $\two h_{ij}$, the gauge conditions (\ref{gauge1})
and the standard post-Newtonian parametrization for matter \cite{will} to write eq.~(\ref{4hoo}) as
\begin{multline}
-\nabla^2(\four h_{00}+2U^2-8{\boldsymbol\Phi_2})=\\\kappa^2 \left[2\rho\left(v^2-U+\frac{1}{2}\Pi-\frac{3p}{2\rho}\right)-V(\phi_0)+\frac12\lambda\dot\phi_0^2\right],
\end{multline}
where $\Pi$ is the specific energy density (ratio of energy density to rest-mass density) \cite{will} and
\be
\Phi_2=G\int d^3 x'\frac{\rho(x',t) U(x',t)}{|x-x'|}.
\ee
The solution to this equation is
\begin{multline}
\four h_{00}=2U^2+4\Phi_1+4\Phi_2+2\Phi_3+6\Phi_4\\+\frac{\kappa^2}{6}\left(V(\phi_0)-\frac12\lambda\dot\phi_0^2\right)|x|^2,
\end{multline}
where
\bea
\Phi_1&=&G\int d^3 x'\frac{\rho(x',t) v(x',t)^2}{|x-x'|},\\
\Phi_3&=&G\int d^3 x'\frac{\rho(x',t)\Pi(x',t) v(x',t)^2}{|x-x'|},\\
\Phi_4&=&G\int d^3 x'\frac{p(x',t)}{|x-x'|}.
\eea
Therefore the metric, expanded in post-Newtonian orders is
\bea
g_{00}&=&-1+2U-2U^2+4\Phi_1+4\Phi_2+\nn\\& &+2\Phi_3+6\Phi_4+\frac{\kappa^2}{6}\left(V(\phi_0)-\frac12\lambda\dot\phi_0^2\right)|x|^2,\\
g_{0j}&=&-\frac{7}{2}V_i-\frac{1}{2}W_i,\\
g_{ij}&=&(1+2U)\delta_{ij},
\eea
which, apart from the term related to $V(\phi_0)-1/2\lambda\dot\phi_0^2$, is exactly the result that one obtains for General Relativity. This term corresponds to the standard correction normally arising from a cosmological constant, and since $V(\phi_0)-1/2\lambda\dot\phi_0^2$ should indeed be of the same order as the energy density of the cosmological constant, the contribution of this term is negligible on Solar system scales. Since the metric is written in the standard PPN gauge one can read off  the PPN parameters \cite{will}. The only non-vanishing ones are $\gamma$ and $\beta$, which are equal to $1$. Therefore, the theory discussed here seems to be indistinguishable from General Relativity at the post-Newtonian order. 

The above implies that a gravitational theory with a scalar field coupled to the Gauss-Bonnet invariant trivially satisfies the constraints imposed on the post-Newtonian parameters by Solar system tests. This appears to be due to the fact that the terms arising in the field equation for the metric from the coupling between the scalar and ${\cal G}$ in the action always have the structure of two derivatives of $f$ times a curvature term. Such terms do not contribute to the post-Newtonian expansion to ${\cal O}(v^4)$. This is not the case for other possible couplings of a scalar to a quadratic curvature term, such as $\phi R^2$. Remarkably, the characteristic structure of such terms can be traced back to the special nature of ${\cal G}$, {\em i.e.~}to the fact that it is a topological invariant in four dimensions.

 However, notice that our result strongly depends on the assumption that $V(\phi_0)$ and $V'(\phi_0)$ are reasonably small so as to give a negligible contribution. This assumption stems from the fact that $V(\phi_0)$ will play the role of an effective cosmological constant if the theory is to account for the late time accelerated expansion of the universe and should therefore be of the relevant order of magnitude. Additionally we expect that $V'(\phi_0)$ will also be small enough so that its contribution can be considered negligible, based on the fact that either the field approaches a minimum at late times, or that the potential is of the form $V=V_0 e^{-a\kappa\phi}$, where $a$ is of order unity, and therefore $V'(\phi_0)\sim\kappa V(\phi_0)$, which seem to be true in all reasonable models.

Another important aspect that has to be stressed is that the value of $f'(\phi_0)$ or $f''(\phi_0)$ should be suitable so that the post-Newtonian expansion remains trustworthy. From eq.~(\ref{phisol}) we see that non-trivial corrections will indeed be present in post-post-Newtonian orders and, even though such corrections are normally subdominant, if $f'(\phi_0)$ or $f''(\phi_0)$ are sufficiently large they can become crucial for the viability of the theory. This was first observed in \cite{farese} where the same theory, but without a potential $V$, was confronted with Solar system observations, considering a nearly Schwarzchild metric as an approximation. As mentioned before, the potential plays the role of an effective cosmological constant if one wants a theory that leads to a late time accelerated expansion as in \cite{nojodsas,cosm}. If this potential is not present, it is the coupling $f(\phi)$ between the scalar field and the Gauss-Bonnet term that will have to account for this phenomenology. In this case it turns out that $f''(\phi_0)$ has to be of the order of the inverse of the cosmological constant, and this is enough to make the post-Newtonian approximation break down. Fortunately, models with a potential do not suffer from this problem, and as a matter of fact $f$ is usually assumed to be of the form $f=f_0 e^{b\kappa \phi}$ where both $f_0$ and $b$ are of order unity. So, as predicted also in \cite{farese}, all reasonable models with a potential will pass the Solar system tests. 

Finally, let us discuss the possibility of including a second scalar field in the action, coupled to the Gauss-Bonnet invariant, which could, for example, be the dilaton. If this second scalar is not coupled to matter or to the Ricci scalar, then it can be treated using the same approach. If the coupling function with the Gauss-Bonnet invariant as well as the potential, if present, have similar properties as those discussed above we expect our result to remain unaffected. Of course there is also the possibility that the dilaton is coupled to matter. This goes beyond the scope of the present paper since in this case the theory would be phenomenologically different not only in what regards Solar system tests but also in other aspect such as cosmology, covariant conservation of matter, equivalence principle ({\em e.g.} see ref.~\cite{taylor}) etc.

\section*{Acknowledgements}
The authors wish to thank Valerio Faraoni, John Miller and Stefano Liberati for a critical reading of this manuscript and constructive suggestions for its improvement.


\end{document}